\documentclass{aa}  
\usepackage{lipsum}
\usepackage{graphicx}
\usepackage{txfonts}	
\usepackage[euler]{textgreek}
\usepackage{cuted}
\usepackage{dblfloatfix}
\usepackage{silence}
\usepackage{amsmath}

\begin{document}
    \title{[Mg/Fe] and variable initial mass function: Revision of [\textalpha/Fe] for massive galaxies}
   
   \subtitle{}
   \author{Emilie Pernet
          \inst{1,2,3}
          \and 
          Alina B\"ocker \inst{1,2,4}
          \and
          Ignacio Mart\'{i}n-Navarro \inst{1,2}}
          
   \institute{Instituto de Astrofísica de Canarias (IAC), E-38200 La Laguna, Tenerife, Spain \\
   \email{emiliepernet.astro@gmail.com}
        \and 
   Departamento de Astrofísica, Universidad de La Laguna, E-38205 La Laguna, Tenerife, Spain
        \and
    Department of Physics, Faculty of Engineering and Physical Sciences, University of Surrey, GU2 7XH, Guildford, United Kingdom
        \and
    Department of Astrophysics, University of Vienna, T\"urkenschanzstrasse 17, 1180 Vienna, Austria}
   \date{Received January 22, 2024; Accepted May 1, 2024}
 
  \abstract{Observations of nearby massive galaxies have revealed that they are older and richer in metals and magnesium than their low-mass counterparts. In particular, the overabundance of magnesium compared to iron, [Mg/Fe], is interpreted to reflect the short star formation history that the current massive galaxies underwent early in the Universe. We present a systematic revision of the [Mg/Fe] - velocity dispersion (\textsigma) relation based on stacked spectra of early-type galaxies with a high signal-to-noise ratio from the Sloan Digital Sky Survey (SDSS). Using the penalized pixel-fitting (pPXF) method of \citealp{Cappellari2004} and the \citealp{Vazdekis2015} MILES single stellar population (SSP) models, we fit a wide optical wavelength range to measure the net \textalpha-abundance. The combination of pPXF and \textalpha-enhanced MILES models incorrectly leads to an apparently decreasing trend of [\textalpha/Fe] with velocity dispersion. We interpret this result as a consequence of variations in the individual abundances of the different \textalpha-elements. This warrants caution for a na\"ive use of full spectral fitting algorithms paired with stellar population models that do not take individual elemental abundance variations into account, especially when deriving averaged quantities such as the mean [$\alpha$/Fe] of a stellar population. In addition, and based on line-strength measurements, we quantify the impact of a non-universal initial mass function on the recovered abundance pattern of galaxies. In particular, we find that a simultaneous fit of the slope of the initial mass function and the [Mg/Fe] results in a shallower [Mg/Fe]-\textsigma\ relation. Therefore, our results suggest that star formation in massive galaxies lasted longer than what has been reported previously, although it still occurred significantly faster than in the solar neighbourhood.}

   \keywords{Galaxies: evolution--
            Galaxies: abundances --
            Galaxies: stellar content --
            Galaxies: elliptical and lenticular, cD
               }

   \maketitle
\nolinenumbers

%__________________________________________________________________
\section{Introduction}

The study of the stellar populations of nearby galaxies has revealed that more massive objects tend to be older and richer in metals. This is a consequence of intense bursts of star formation that occurred at early times \citep[e.g.][]{2_Trager2000}. The analysis of their spectra through absorption lines uncovered a strong correlation between the strength of magnesium absorption features and the galaxy luminosity or velocity dispersion ($\sigma$; \citealp{Peletier1989, Worthey1992}). This phenomenon was found to be linked to the previously mentioned increase in the age and metallicity, but also to the increase in [Mg/Fe] as a function of $\sigma$ \citep{Thomas2005}.

The reason for the Mg-enhancement goes back to the nucleosynthesis of magnesium-like elements and iron-peak elements, which are released through different mechanisms: core-collapse and type Ia supernova (SN) explosions, respectively \citep{Thielemann2003}. Three explanations were proposed to account for the increase in [Mg/Fe], often generalised as [\textalpha/Fe], in massive early-type galaxies
%\LEt{***please introduce all your abbreviations and acronyms at first occurrence in the maint text (because the Abstract is considered a stand-alone item, and when an abbreviation is used fewer than three times in the Abstract or fewer than five times in the main text, they remain spelled out) and then use them without reintroduction throughout for consistency, including all figure captions and tables. Please check this and change throughout as required. I'll not highlight this again to avoid cluttering the ms***} 
(ETGs) \citep{Worthey1992, Faber1992}: (1) selective mass loss, with SN-driven winds causing strong loss of Mg-like elements in less massive galaxies, (2) a non-universal initial mass function (IMF), allowing massive galaxies to have a greater number of massive stars, resulting in the release of larger amounts of Mg, or (3) a different star formation timescale that is short and intense for massive galaxies. The latter was established as the favoured explanation because models of selective mass-loss mechanisms have an opposite impact \citep{Matteucci1994}, and observations of nearby systems suggested a universal Milky Way-like IMF \citep{Kroupa2001, Chabrier2003}, so that the variable IMF hypothesis was discarded.

New observational and theoretical advances have recently enriched our view of the \textalpha-enhancement in galaxies, however. In particular, observations have demonstrated that a universal IMF is inconsistent with observations of massive ETGs because their central regions seem to exhibit an enhanced fraction of low-mass stars \citep{Conroy2012, Cappellari2012, LaBarbera2013}. In addition, through the advent of the latest generation of integral field spectroscopic units  \citep[e.g.][]{Bacon2017}, it is now possible to conduct a detailed stellar population analysis across entire galaxies, which can unveil the two-dimensional spatial variation in [Mg/Fe] \citep[e.g.][]{Pinna2019b, Pinna2019, Martin2021}. Finally, the most advanced cosmological numerical simulations begin to accurately track the evolution of \textalpha-elements, setting direct constraints on the black hole feedback implementation  \citep[e.g.][]{Segers2016} and on the universality of the IMF \citep[e.g.][]{Bekki2013}.

Motivated by these recent developments, we re-assess here the robustness of [\textalpha/Fe] measurements from integrated spectra using the latest data, models, and analysis tools. We show that in the context of a full spectral fitting (FSF), measuring [\textalpha/Fe] can lead to nonphysical results when individual elemental variations are neglected. We also show that a variable IMF has a non-negligible effect on the recovered [Mg/Fe] ratio. The layout of this paper is as follows: in \S~\ref{sec:data}, we describe the data and stellar population modes. In \S~\ref{sec:develop}, we describe the results we obtained using pPXF, and in \S~\ref{sec:fif}, we describe the results that are based on a line-strength analysis to assess the impact of a variable IMF. Finally, in \S~\ref{sec:end}, we discuss and summarise the main implications of our measurements.

%__________________________________________________________________
\section{Data and stellar population models}\label{sec:data}

We analysed 18 stacked spectra from the
%\LEt{***please provide the spelled-out versions of all instruments and surveys at first occurrence. This may be done as a footnote if it interrupts the flow of the sentence too much. Please check this throughout and change as required***}
SPIDER \footnote{SPectroscopic IDentfication of ERosita Sources} sample assembled by \citet{LaBarbera2013}. Each stacked spectrum consists of hundreds of individual spectra of ETGs with a similar velocity dispersion taken from the Sloan Digital Sky Survey (SDSS) DR6 \citep{Adelman-Mccarthy2008}, ranging from 100 to 320 $km\,s^{-1}$. The stacked spectra reach high signal-to-noise ratios, which makes this sample ideal to perform detailed stellar population analysis.

Unless stated otherwise, we based our analysis on the MILES single stellar population (SSP) models (\citealp{Vazdekis2010}; \citealp[updated in][]{Vazdekis2015}), which cover the full 3800-7500\,\AA\,wavelength range with a resolution of 2.51 \,\AA\, FWHM \citep{Falcón-Barroso2011}. We selected the models generated with BaSTI isochrones \citep{Pietrinferni2004, Pietrinferni2006}, with ages from $0.9$ to $14$ Gyr,  metallicities [M/H] from $-2.27$ to $+0.40$ dex, and two [\textalpha/Fe] predictions, $+0.0$ and $+0.4$ dex. It is worth highlighting that in these models, all \textalpha-elements were varied in lockstep \citep[see][Section 3.2]{Vazdekis2015}. We complemented the MILES models with the  \citet{Conroy2012_models, Conroy2014} predictions to assess the effect of varying individual abundances (specifically, C, Mg, Ca, Ti, and Si) in the optical spectra. 

%__________________________________________________________________
\section{Measuring [\textalpha/Fe] with the MILES models and pPXF} \label{sec:develop}

We fed the MILES models described previously into pPXF (\citealp{Cappellari2004}; \citealp[updated in][]{Cappellari2017}). In short, pPXF is an inversion algorithm that finds the linear combination of SSPs to best represent an observed spectrum. We fit a wavelength range of 4000-6600\,\AA, including a multiplicative Legendre polynomial of order 13 to correct for continuum mismatch between models and SDSS data. We fixed the regularisation parameter to $reg = 1/0.5$ because we are only interested in average stellar population quantities for this study \citep{Boecker2020}. To measure errors on stellar population properties, we performed Monte Carlo simulations by generating random Gaussian noise, scaled to the flux uncertainties of each spectrum. The stellar population properties recovered in this way are mass-weighted averages by construction since MILES models are normalised to one solar mass.

%__________________________________________________________________
\subsection{Misleading [\textalpha/Fe]-\textsigma\ trends} \label{sec:fsf}

\begin{figure*}
    \begin{center}
    \includegraphics[scale=0.17]{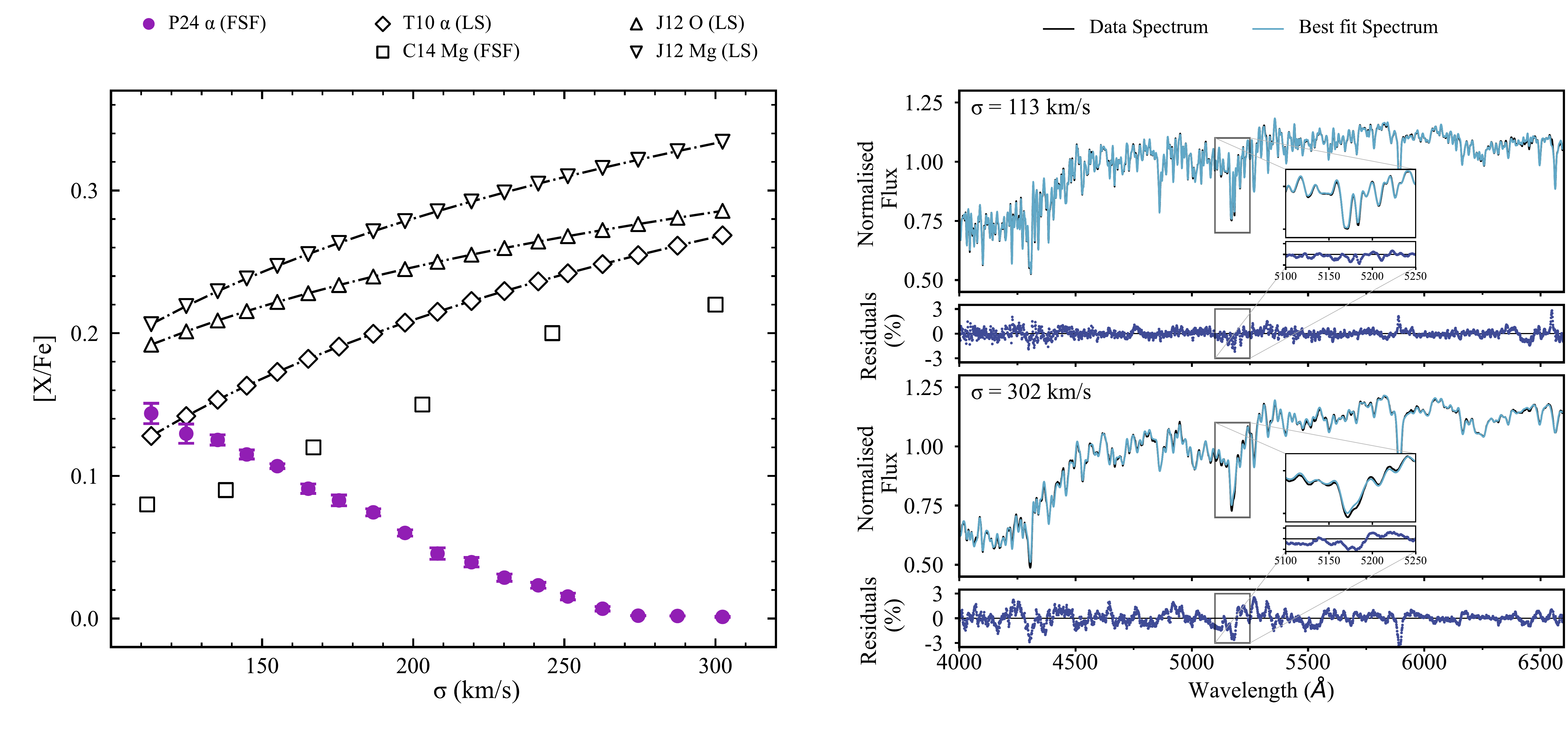}
    \caption{Measure of [$\alpha$/Fe] on SDSS stacked spectra and associated best-fit spectra. Left 
    %\LEt{***please add a short descriptive sentence omitting the initial article before you describe the individual panels. This applies to all figure captions. Please check and change as required.***}
    panel: Elemental abundances as a function of the velocity dispersion for ETGs. Empty symbols correspond to literature measurements of different $\alpha$ elements. The diamonds indicate average [$\alpha$/Fe] from \citet{Thomas2010}, up and down triangles are oxygen and magnesium as measured by \cite{Johansson2012}, and squares show the [Mg/Fe] trend of \cite{Conroy2014}. The filled purple circles show the best-fitting (but misleading) pPXF+MILES solution. Right panels: Spectra for the lowest (top panel) and highest (bottom panel) \textsigma\ bin. The observed stacked spectrum is given in black and the best-fit spectrum in cyan. The residuals are shown in blue. We also provide a zoom-in into the Mgb region.}
    \label{fig:fig1}
    \end{center}
\end{figure*}

In Figure \ref{fig:fig1} (left panel) we show the recovered [\textalpha/Fe] values of all 18 stacked spectra using the setup (pPXF+MILES) described above as a function of their velocity dispersion from our FSF analysis. In contrast to previous results, we recover an apparent decrease in [\textalpha/Fe] for galaxies with a higher velocity dispersion. For comparison, Figure \ref{fig:fig1} also shows the literature values from \cite{Thomas2010}\footnote{Given the set of indices included in the analysis of \citep{Thomas2010} it is fair to assume that [\textalpha/Fe] in this case corresponds to [Mg/Fe].}, \cite{Johansson2012}, and \cite{Conroy2014}. While there are obvious systematic differences among these measurements, they all agree qualitatively on the well-established trend of [\textalpha/Fe] with \textsigma. While the apparent [\textalpha/Fe]-\textsigma\, trend recovered here contrasts with previous works, the age and metallicity trends increase with \textsigma, which agrees with the literature.

To assess the origin of this misleading [\textalpha/Fe] trend that results from combining pPXF with the MILES models, we show the best-fit spectra for the lowest and highest $\sigma$ on the right side of Figure \ref{fig:fig1}. Even though the residuals are lower than 3\%, the magnesium triplet (5175\AA) is clearly not well fit by the models, especially for the high $\sigma$ bin. This suggests that the [\textalpha/Fe] obtained with the MILES models and pPXF do not capture the actual change in [Mg/Fe].

\begin{figure}[!hbt]
    \begin{center}
    \includegraphics[width=8.5cm]{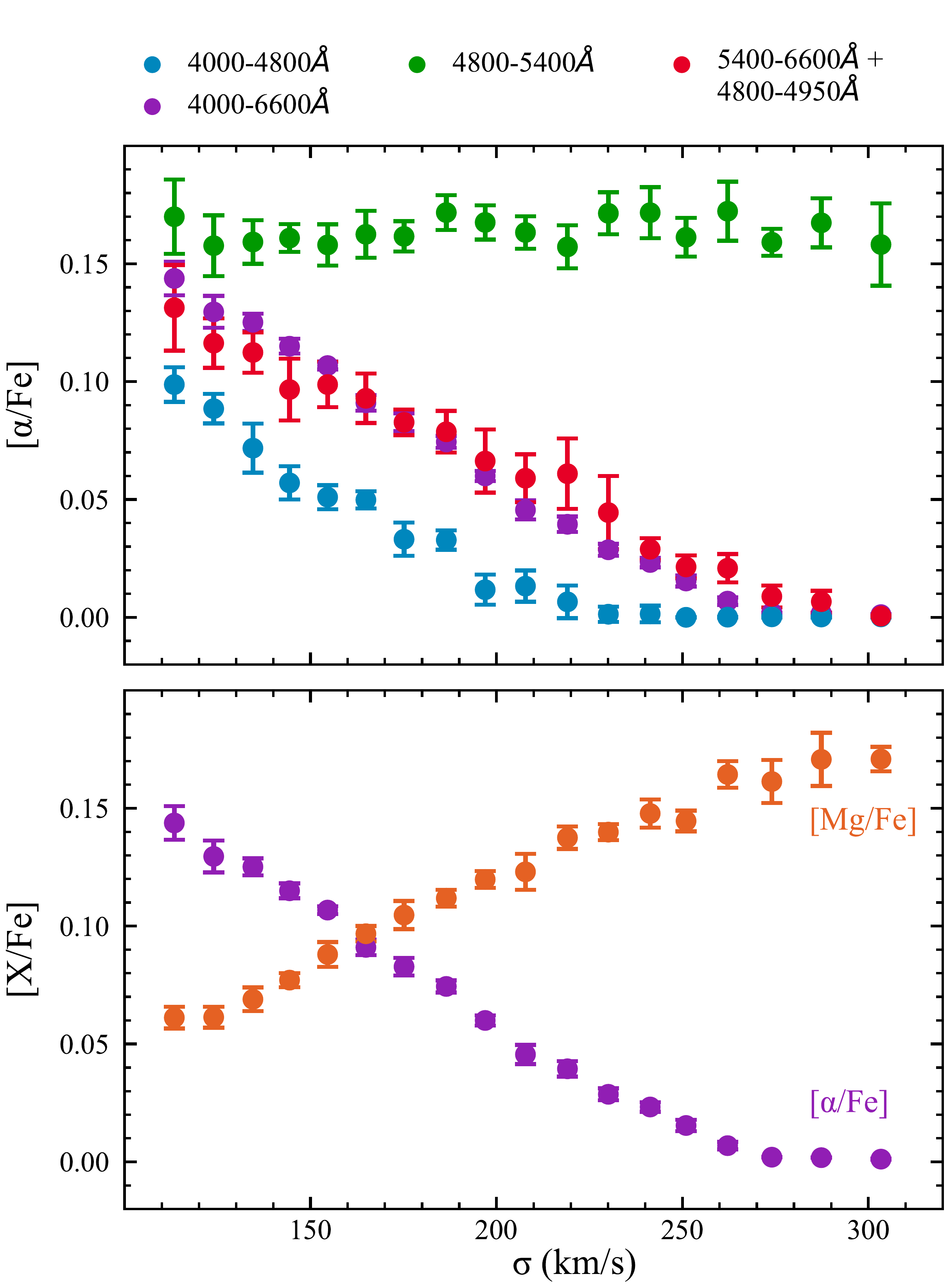}
    \caption{Further measures of [$\alpha$/Fe], and [Mg/Fe]. Top panel: [\textalpha/Fe]-\textsigma\ relation obtained by fitting the SDSS data over different wavelength ranges: 4000-4800\,\AA\,(blue), 4800-5400\,\AA\,(green), and 5400-6600\,\AA\,including the H$\beta$ region (red). Bottom panel: [X/Fe]-\textsigma\ relation using magnesium-enhanced models (orange). These models combine the MILES \citet{Vazdekis2015} and the \citet{Conroy2012_models} predictions (see details in the text). For reference, the filled purple circles in both panels represent the results from Figure \ref{fig:fig1}.}
    \label{fig:fig2}
    \end{center}
\end{figure}

As an additional test, we repeated our stellar population analysis over three different wavelength regions: 1) 4000-4800\,\AA, 2) 4800-5400\,\AA\,and 3) 5400-6600\,\AA\,including the H$\beta$ region. We display the results for [\textalpha/Fe] in the top panel of Figure \ref{fig:fig2} and compare them to our previous results.
We obtain similar results, that is, an apparent decrease in [\textalpha/Fe] with increasing \textsigma, for the bluest and reddest wavelength regions. Only for the region dominated by the Mgb triplet (green symbols), are we able to recover a flat trend between the measured [\textalpha/Fe] and \textsigma.

%__________________________________________________________________
\subsection{Individual \textalpha-element abundances} \label{sec:fsfelem}

The tests above, and in particular, the right panels in Figure \ref{fig:fig1}, reveal that the combination of \textalpha-enhanced MILES models and pPXF cannot model the strength of the Mgb absorption feature properly, which might lead to biased [\textalpha/Fe] measurements. We only recover a non-decreasing trend when we focus on the magnesium triplet. Aiming to isolate the effect of [Mg/Fe] on the recovered trends, we repeated our stellar population analysis with a set of magnesium-enhanced models. These models were constructed by combining MILES base models (i.e. with an abundance pattern that traces the [Mg/Fe]-[Fe/H] relation of the solar neighbourhood) with the magnesium response function obtained by \cite{Conroy2012_models}. After feeding pPXF with these Mg-enhanced models over the entire wavelength range 4000-6600\,\AA, we obtained the trend shown in the bottom panel of Figure \ref{fig:fig2}.   

The bottom panel of Figure \ref{fig:fig2} clearly shows that when fed with Mg-only enhanced models, pPXF is able to recover the expected trend between stellar velocity dispersion and [Mg/Fe] (orange symbols), and it only fails when the input models assume that all \textalpha-elements are enhanced (purple symbols). Since in the \textalpha-variable MILES models all \textalpha-elements are varied in lockstep, it is likely that the trend shown in Figure \ref{fig:fig1} is caused by the failure of the different \textalpha-elements to track each other as a function of velocity dispersion \citep[e.g., ][]{Conroy2014}. We also find similarly spurious trends when using the different predictions of \citet{Conroy2012_models} to compute \textalpha-variable models, but with fixed relative abundances (e.g. all \textalpha-elements are enhanced to the same degree).

To test this hypothesis, we emulated the effect that changes in the mixture of \textalpha-elements would have on the observed SDSS data. We constructed SDSS-like mock spectra using the MILES [\textalpha/Fe]=0 models and corrected them with the \citet{Conroy2012_models} predictions according to the individual abundance ratios found by \citet{Beverage2023} for C, Mg, Si, Ca, and Ti. This allowed us to explore the effect of measuring [\textalpha/Fe] with SSP models that enhance all \textalpha-elements simultaneously by the same amount on data for which the elements do not follow this behaviour. Over these mock spectra, we then ran our fiducial stellar population analysis, that is, pPXF plus the \textalpha-variable MILES models. 

\begin{figure}[!hbt]
    \begin{center}
    \includegraphics[width=8.5cm]{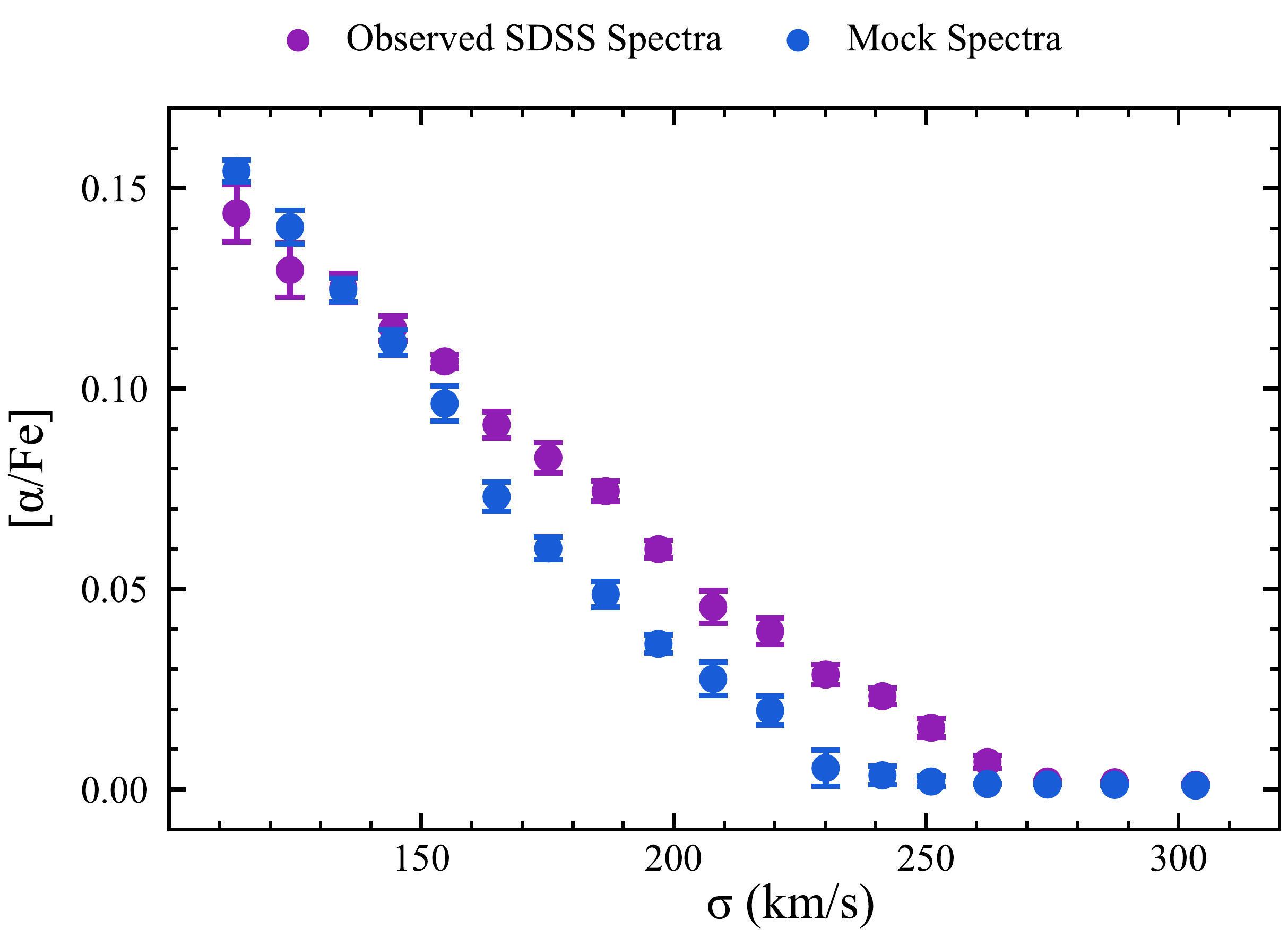}
    \caption{ [\textalpha/Fe]-\textsigma\ relation measured from the mock spectra. The blue circles represent the results obtained for the mock spectra that include different abundance patterns for individual\textalpha-elements as measured in \citealp{Beverage2023}; \citealp[updated results from][]{Conroy2014}. The filled purple circles again represent the results found in Figure \ref{fig:fig1} for comparison purposes.}
    \label{fig:fig3}
    \end{center}
\end{figure} 

The result of this test is shown in Fig.~\ref{fig:fig3}. We recover a remarkably similar decreasing trend of [\textalpha/Fe] from the mock spectra (blue symbols) compared to our initial results of the real SDSS data (purple symbols). This suggests that applying models that vary all \textalpha-elements by the same amount to data for which different \textalpha-elements do not track each other might lead to biased and thus flawed measurements. 

Our understanding of this phenomenon relies on the non-uniformity of flux variations caused by the enhancement of individual \textalpha-elements. Essentially, an increase in the abundance of individual \textalpha-elements can yield deeper absorption features for some elements while weakening them for others \citet[see][Figure 2]{Conroy2014}. Therefore, this could result in flux changes that compensate for each other to some extent within specific wavelength regions, even when all these elements are varied by the same amount. This ill-constrained behaviour is further accentuated by the fact that the resulting effective [\textalpha/Fe] is weighted by the wavelength-dependent sensitivity to the elemental abundance. Hence, applying a model in which all \textalpha-elements are varied in lockstep to a spectrum in which individual \textalpha-elements likely have different abundances could result in the FSF trying to fit certain wavelength regions in which the flux change of all the \textalpha-elements combined translates into a decreasing [\textalpha/Fe] trend. However, it is important to note that all individual \textalpha-elements do not truly decrease due to the different relative flux changes described above.
%_____________________________
\section{Effect of a variable initial mass function on the [Mg/Fe]-\textsigma\ relation} \label{sec:fif}

The results above highlight some of the biases that can occur from an incorrect application of SSP models that enhance all \textalpha-elements in the same way, in combination with a large wavelength range, as is the case for a full spectral fitting. This is particularly the case for information that is predominantly concentrated on specific absorption features. In this section, we make use of the high sensitivity of some of these absorption features to revisit the [Mg/Fe]-\textsigma\ relation in the context of a non-universal IMF. 

We followed the full index-fitting (FIF) approach described in \citet{Martin2019, Martin2021} to measure the age, metallicity, [Mg/Fe], IMF slope, and [Ti/Fe] of the stacked SDSS spectra. This fitting approach consists of a Bayesian fit to each pixel within the (continuum-corrected) bandpass definition of a set of key spectral features, which in this case were H$\beta_o$, an optimised version of H$\beta$ \citep{Cervantes2008}, Mgb, Fe\,5015, Fe\,5270, Fe\,5335, TiO$_1$, and TiO$_2$. We used MILES \textalpha-enhanced SSP models with a bimodal IMF slope \citep{1996vazdekis} ranging from $\Gamma_{b}= 0.5$ to $\Gamma_{b}= 3.5$. However, we measured [Mg/Fe] because of the indices we fitted \citep{Martin2019, Martin2021}. In contrast to our previous measurements using pPXF, FIF-based quantities are weighted by luminosity\footnote{pPXF fits a linear combination of multiple SSPs to the observed spectrum, whereas FIF finds the best-fit SSP-equivalent spectrum.}.

\begin{figure*}
    \begin{center}
    \includegraphics[scale=0.21]{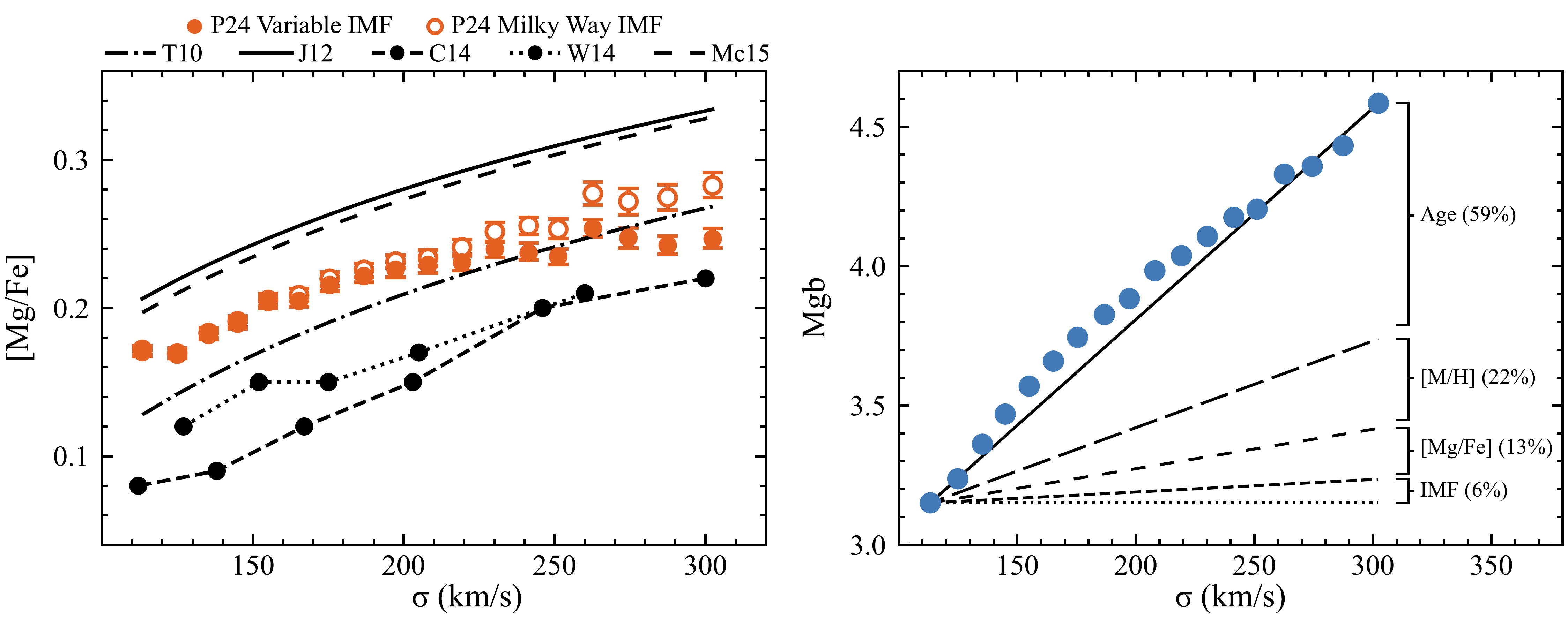}
    \caption{The impact of the IMF on [Mg/Fe] and Mgb measurments. Left panel: [Mg/Fe]-\textsigma\ relation of our stack spectra using FIF (orange). The empty circles represent the results obtained with a fixed Milky Way-like IMF, and the filled circles show our measurements when the IMF is treated as an additional free parameter. The black lines correspond to literature [Mg/Fe] measurements: \citet{Johansson2012} (solid line), \citet{Worthey2014} (dotted line with data points), \citet{Conroy2014} (dashed line with data points), \citet{Thomas2005} (dash-dotted line), and \citet{McDermid2015} (dashed line). Right panel: Mgb-\textsigma\ relation of our stack spectra and the relative contribution to the Mgb strength from age, metallicity, [Mg/Fe], and IMF variations.}
    \label{fig:fig4}
    \end{center}
\end{figure*}

The left panel of Fig.~\ref{fig:fig4} shows our best-fitting [Mg/Fe] values as a function of the stellar velocity dispersion. The measurements allowing for a variable IMF in the fitting process are shown as filled orange symbols, and a fixed Milky Way-like IMF is shown as empty orange symbols. Our measurements based on a universal IMF agree with the literature, as expected. We measure an increase of 0.11 dex in [Mg/Fe] from the lowest to the highest velocity dispersion stacks. A similar trend emerges from the variable IMF results, up to \textsigma\ $\sim$ 230 $km s^{-1}$. However, for \textsigma\ $\gtrsim$ 240 $km s^{-1}$, the trend flattens, resulting in lower [Mg/Fe] for the most massive galaxies. This is ultimately due to the non-negligible dependence of the Mgb absorption feature on the adopted IMF within the MILES models. The IMF slopes retrieved with FIF perfectly agree with the results found in \citet{LaBarbera2013}. Regardless of the assumptions on the IMF, [Mg/Fe] clearly increases with \textsigma, which emphasises the spurious nature of the trends shown in Fig.~\ref{fig:fig1}.

To further illustrate the importance of a variable IMF in our interpretation of the [Mg/Fe] enhancements of massive galaxies, the right panel of Fig.~\ref{fig:fig4} shows the age, metallicity, [Mg/Fe], and IMF variations to the empirical Mgb-$\sigma$ relation. As reported by \citet{Thomas2005}, this relation encodes the fact that more massive ETGs are older, more metal rich, and more enhanced in [Mg/Fe] than their low-mass counterparts. However, there is a subtle but noticeable additional contribution to the Mgb strength from the excess of low-mass stars that are contained in the central regions of massive ETGs. 

%______________________________
\section{Discussion and conclusion} \label{sec:end}

We have demonstrated that the measure of [\textalpha/Fe] as a function of \textsigma\ using pPXF and the MILES SSP models over a wide wavelength range results in an apparently decreasing trend. Even when a narrower wavelength range is selected that focused on the Mgb region, we still did not recover the well-known increasing trend. Moreover, we find significant differences in the recovered trends based on the \citet{Vazdekis2015} or \citet{Conroy2012_models} models. Both models lead to spurious results when fed into pPXF, however. These results do not reflect an actual decrease in [\textalpha/Fe] for more massive ETGs, but the limitations of FSF algorithms combined with SSP models assuming a fixed mixture of \textalpha-elements to capture the complexity of the chemical evolution in galaxies. Discrepancies among the model predictions further emphasise the difficulty of robustly quantifying a single and representative [\textalpha/Fe] value.

Our results therefore call for a careful interpretation of [\textalpha/Fe] measurements based on the combination of FSF tools and SSP models that only vary the global [\textalpha/Fe] value. Since not all \textalpha-elements necessarily track each other \citep[e.g.][]{Worthey1998, Johansson2012, Conroy2014, Worthey2014}, the same [\textalpha/Fe] can result from several completely different combinations of individual abundances. Furthermore, as revealed by our analysis, [\textalpha/Fe] estimates might be completely unphysical when the assumed stellar population models do not allow for relative changes in elemental abundances. This is particularly problematic with FSF algorithms because different \textalpha-elements contribute differently and in a highly degenerate manner to absorption spectra of galaxies.  

In this context, measurements of individual abundances are more robust quantities, allowing fairer comparisons and a more accurate description of the underlying physical mechanisms controlling the evolution of galaxies. For example, the measure of magnesium is the founding principle of what is now qualified as \textalpha-enhancement \citep{Peletier1989, Worthey1992}. In addition, light \textalpha-elements, such as Mg or Si, are more strongly enhanced than heavier \textalpha-elements, such as Ca and Ti \citep{Johansson2012, Conroy2014, Worthey2014}. Furthermore, Milky Way abundance studies revealed that magnesium is the most robust and consistent \textalpha-element \citep{Jonsson2018, Zasowski2019}. Therefore, [Mg/Fe] alone acts as a better tracer than [\textalpha/Fe], as expected.

We also showed that precise [Mg/Fe] are sensitive to IMF variations, in agreement with \citet{Conroy2014}. For the most massive galaxies in the SDSS stacked spectra, which exhibit the largest IMF variations, we find [Mg/Fe] values that are lower by $\sim$0.035 dex than our measurements assuming a Milky Way-like IMF. Under the assumption that [Mg/Fe] scales with the timescales of star formation \citep[e.g.][]{Thomas2005}, our results then indicate that massive ETGs formed their stellar component over relatively longer periods of time than is usually assumed \citep[e.g.][]{McDermid2015,Segers2016}.

These findings imply that our physical interpretation of the \textalpha-enhancement as only reflecting the length of the star formation timescales is not complete and among other aspects, lacks the implications of a varying IMF. A more nuanced interpretation that combines the effect of star formation and the non-universal IMF will be necessary to be better represent the observations. The consideration of a time-varying or a non-canonical IMF will open the door for a more complete and consistent characterisation of the chemical evolution of massive galaxies \citep[e.g.][]{1997Vazdekis,2015Ferreras,2018Tereza}

\begin{acknowledgements}
We would like to thank the anonymous referee for their detailed comments and suggestions, resulting in an improved version of the manuscript. EP would like to thank Francesco La Barbera for his stacked spectra and insights, and Alexandre Vazdekis for his support and enriching discussions. EP acknowledges support from the Turning Scheme. We acknowledge support from grants PID2019-107427GB-C32 and PID2022-140869NB-I00 from the Spanish Ministry of Science. AB gratefully acknowledges support from the Moritz-Schlick early-career Postdoc Programme.
\end{acknowledgements}

\bibliographystyle{aa} 
\bibliography{References} 

\begin{appendix}
\section{Variations in the initial mass function of FIF and FSF}\label{appB}

\begin{figure}[!hbt]
    \includegraphics[width=\columnwidth]{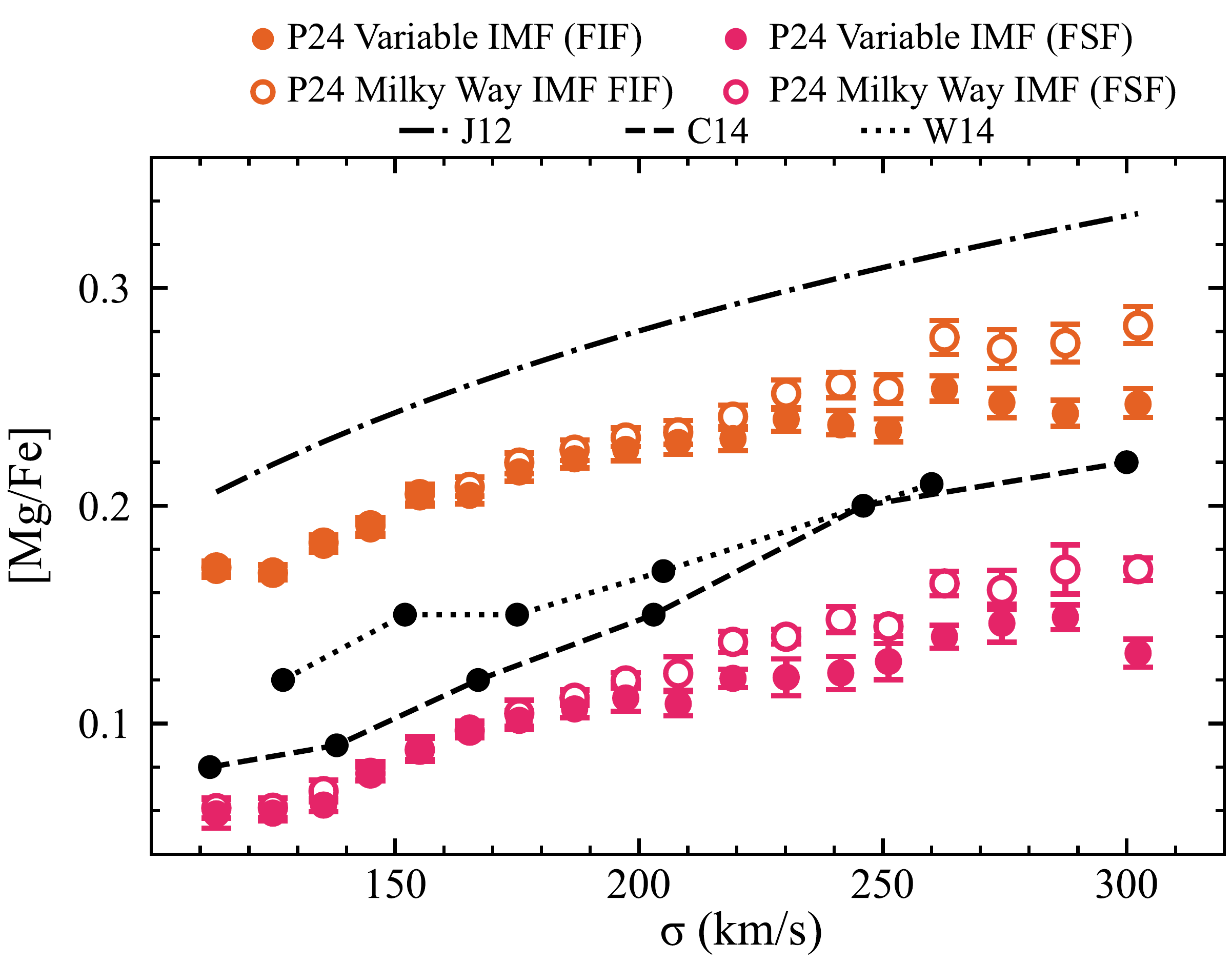}
    \caption{[Mg/Fe]-\textsigma\ relation of our stack spectra measured using FSF and Mg-enhanced models (pink) as well as FIF (orange) for a Kroupa universal IMF (empty circles) and a $\sigma$-varying IMF (filled circles). The black lines and symbols correspond to works (also shown in Figure \ref{fig:fig4}) that explicitly measured [Mg/Fe].}
    \label{fig:fig6}
    
\end{figure}

We show the [Mg/Fe]-\textsigma\ using FIF and FSF with both a fixed and variable IMF in Figure \ref{fig:fig6}. 

The results obtained with the two methods agree with each other. The difference between the results from FSF and FIF corresponds to a shift of $\sim$ 0.1 dex. The same shift is observed between \citet{Johansson2012} and \citet{Conroy2014}. These results show that even with FSF, using a wide wavelength range and SSP models that are only enhanced in magnesium, the impact of the IMF can be measured. Moreover, this provides an additional piece of evidence that [Mg/Fe] is a robust \textalpha-element.

\end{appendix}
\end{document}